\documentclass[runningheads]{llncs}
\usepackage{graphicx}
\usepackage[nolist]{acronym}
\usepackage[caption=false]{subfig}
\usepackage{multirow}
\usepackage{xcolor}

\usepackage{todonotes}
\usepackage[hidelinks]{hyperref}
\usepackage{ulem}  

\begin{document}

\newcommand\copyrighttext{ 
	\noindent\Huge {EAI Copyright Notice} \\ \\
	\noindent\large {\textcopyright~2021 EAI. Authors have the right to post and update author-prepared versions of the article on free-access e-print servers, including the author and/or employer’s home page and any repository legally mandated by the agency funding the research on which the Work is based. If the author wishes the EAI-prepared version to be used for an online posting, permission is required from EAI; if granted, use will be subject to EAI terms and conditions.} \\ \\

	\noindent Accepted to be published in:
	\begin{itemize}
		\item Proceedings of 13th EAI International Conference on Simulation Tools and Techniques (EAI SIMUtools 2021), Nov 5-6, 2021.
	\end{itemize}
}


{
\clearpage\thispagestyle{empty}\addtocounter{page}{-1}
\copyrighttext
}

\title{Native versus Overlay-based NDN over Wi-Fi 6 for the Internet of Vehicles}
%
%

\author{Ygor Amaral B. L. de Sena\inst{1,2} \and Kelvin Lopes Dias\inst{1}}

\authorrunning{Y. A. B. L. de Sena et al.}

%

\institute{Centro de Informática, Universidade Federal de Pernambuco, Recife, Brazil\\
\email{ygor.amaral@ufrpe.br, kld@cin.ufpe.br} \and Unidade Acadêmica de Serra Talhada, Universidade Federal Rural de Pernambuco, Serra Talhada, Brazil}

\begin{acronym}[ACRONYM] 

\acro{CS}[CS]{Content Store}
\acro{GI}[GI]{Guard Interval}
\acro{GPS}[GPS]{Global Positioning System}
\acro{IoV}[IoV]{Internet of Vehicles}
\acro{MAC}[MAC]{Medium Access Control}
\acro{MCS}[MCS]{Modulation and Coding Set}
\acro{NDN}[NDN]{Named Data Networking}
\acro{NFD}[NFD]{NDN Forwarding Daemon}
\acro{NFV}[NFV]{Network Function Virtualization}
\acro{PIT}[PIT]{Pending Interest Table}
\acro{SUMO}[SUMO]{Simulation of Urban Mobility}
\acro{VNDN}[VNDN]{Vehicular NDN}
\acro{WLAN}[WLAN]{Wireless Local Area Network}
\end{acronym}

\maketitle              
\begin{abstract}
Internet of Vehicles (IoV) is a cornerstone building block of smart cities to provide better traffic safety and mobile infotainment. Recently, improved efficiency in WLAN-based dense scenarios has become widespread through Wi-Fi 6, a license-free spectrum technology that can complement the cellular-based infrastructure for IoV. In addition, Named Data Networking (NDN) is a promising Internet architecture to accomplish content distribution in dynamic IoV scenarios. However, NDN deployments, i.e., native (clean-slate) and overlay (running on top of IP stack), require further investigation of their performance over wireless networks, particularly regarding the IoV scenario. This paper performs a comparative simulation-based study of these NDN deployments over Wi-Fi 6 for IoV using real vehicular traces. To the best of our knowledge, this is the first effort that extends ndnSIM 2 with an overlay-based NDN implementation and that compares it with the native approach. Results show that the overlay-based NDN consistently outperforms the native one, reaching around 99\% of requests satisfied, against only 42.35\% in the best case of native deployment.

\end{abstract}

\section{Introduction}\label{sec:introduction}
Vehicular Networks have attracted much attention from the industry and research community since an increasing number of vehicles will be connected to the wireless infrastructure of smart cities. Besides traditional traffic safety applications, bandwidth-hungry infotainment services will require efficient distribution of content to vehicles. Despite the wide-area coverage of cellular communications and 5G advancements for vehicular networking, access via \ac{WLAN} has also evolved as an outdoor alternative for free of charge / public access and complementary technology to cellular connectivity. Nowadays, the IEEE 802.11ax standard, also known as Wi-Fi 6, promotes highly efficient communication in dense scenarios~\cite{Khorov2019}. This way, \ac{WLAN} connectivity has gained momentum in smart cities. Similar to cellular networks, the IEEE 802.11 family has evolved and can benefit the efficient content distribution within limited coverage for mobile users and vehicles.

Recently, \ac{NDN}~\cite{Zhang2014}, a promising Internet architecture for content distribution, has also embraced the realm of vehicular networks through the so-called \ac{VNDN}~\cite{Grassi2014}. Instead of using the traditional end-to-end IP-based communication, \ac{NDN} adopts a hop-by-hop approach to distributing and retrieving content on the Internet. Thus, \ac{NDN} does not need network layer addressing but relies on names to request the desired content. This solution has several advantages, especially in mobility contexts. When it comes to vehicles as end-users, this architecture has been promoted to overcome the intrinsic dynamic and challenging scenarios of wireless networks and, in particular, is well-suited to the \ac{IoV} through different solutions based on \ac{VNDN}.

However, \ac{NDN} is a clean-slate network architecture, breaking the compatibility with applications devised to provide services over IP networks. We refer to this approach as native \ac{NDN}, because it performs a full replacement of IP protocol and a direct \ac{NDN} deployment over the link layer~\cite{Nour2019}. In order to tackle this issue, the \ac{NDN} can also be deployed as an overlay network, that is, \ac{NDN} over IP~\cite{Zhang2014,Afanasyev2018a}, thus, enabling coexistence with the traditional Internet architecture and benefiting from existing IP-based applications. We call this deployment overlay \ac{NDN}. In general, \ac{NDN} networks in testbed experiments commonly use only the overlay \ac{NDN} deployment, while proposals evaluated through simulated environments commonly use only native \ac{NDN}.

Despite existing works on the synergy between \ac{NDN} and \ac{IoV}~\cite{Grassi2014,Anastasiades2016,Coutinho2018,Duarte2019,Wang2020}, they analyzed their proposal either considering native or overlay deployment, but there are no insights on the comparative performance of both approaches.
This is due to the lack of an out-of-the-box tool to evaluate both deployments. Furthermore, when modeling vehicle traffic, the works do not consider real traces from transportation authority and use low-throughput wireless networks, which limits the comparison between different approaches. Besides that, contention-based wireless technologies such as Wi-Fi 6 may suffer from broadcast storms or degradation of transmission rates due to the switching to basic service~\cite{IEEE802.11-2021}.

Hence, we conducted a performance evaluation of native and overlay deployments in an \ac{IoV} context considering a Wi-Fi 6 hotspot. The vehicular traffic has been modeled based on real traces. To the best of our knowledge, this is the first effort to implement the overlay \ac{NDN} in ndnSIM 2~\cite{Mastorakis2017a} and perform a comparative evaluation with native deployment. Our simulation results show that the overlay deployment outperforms native in our scenarios, always reaching around 99\% of requests satisfied, against only 42.35\% in the best case of native deployment.

The remainder of this paper is organized as follows. Section~\ref{sec:ndn} describes the general architecture of \ac{NDN} to support different deployments flavors. In Section~\ref{sec:related_work} we describe related work. We address all the implementation details of the simulation-based experiments in Section~\ref{sec:experimental_setup}. We then discuss the performance evaluation results in Section~\ref{sec:performance_evaluation}. Finally, we conclude our paper in Section~\ref{sec:conclusions}.

\section{NDN Deployment Flavors}\label{sec:ndn}
\begin{figure}[htbp]
	\centerline{\includegraphics[width=0.8\columnwidth]{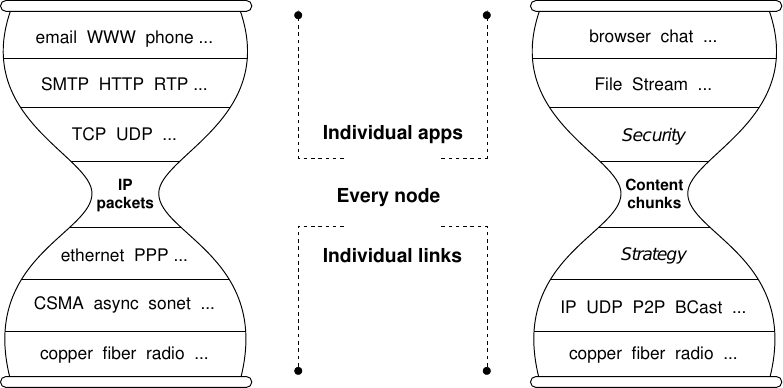}}
	\caption{Architectural differences between IP and \ac{NDN} networks~\cite{Zhang2014}.}
	\label{fig:ndn}
\end{figure}

\ac{NDN} networks have an architecture that works differently than IP networks, focusing mainly on content rather than device addressing, as shown in the narrow waist in Figure~\ref{fig:ndn}. The problem with addressing IP networks is that they rely on end-to-end application communications and depend on the device's location. With the emergence of \ac{IoV}, IP addressing has become a limitation due to the high mobility of these nodes.

Applications on \ac{NDN} networks request data by using the name of desired content, without informing the destination address, through hop-by-hop communication. This modus operandi brings several advantages. One of them is that the \ac{NDN} architecture remains simple, and despite being clean-slate, \ac{NDN} is also a universal overlay; it can run as an overlay network over anything that can forward packets, including IP networks~\cite{Zhang2014}, as also shown in Figure~\ref{fig:ndn}.

Even when the \ac{NDN} is deployed as an overlay network over IP, hop-by-hop communication is employed if there are intermediate NDN nodes between consumer and producer and thus the change of IP address due to mobility should not have the same negative impact when compared to a purely IP-based network. In this way, the overlay \ac{NDN} provides its primary feature of distributing named-based content, while taking advantage of the existing and mature infrastructure of IP networks. However, this deployment increases the network management complexity due to the extra layer of tunnel handling that encapsulates \ac{NDN} packets~\cite{Nour2019}.

Although native deployment is the goal preconized by \ac{NDN} researchers, IP networks are ubiquitous. Therefore, \ac{NDN} needs incremental penetration on the Internet, either on top of IP networks or linked via translators, to be feasible in current real-world scenarios. As described earlier, \ac{NDN} is a universal overlay and runs over IP. In addition, approaches that interconnect native \ac{NDN} and IP networks through translator gateways have been proposed~\cite{Fahrianto2021a,Nour2019}, where IP nodes can communicate with \ac{NDN} nodes and the other way around, without the need for overlapping. In this way, these network architectures must coexist, and after reaching a sufficient number of \ac{NDN} nodes, the IP protocol could be gradually replaced by native \ac{NDN}. Thus, the native deployment must perform satisfactorily during a transition period.

The \ac{NDN} project provides its tools with an open-source license, including \ac{NFD}~\cite{Afanasyev2018a}. This official forwarder has been designed to support the \ac{NDN} in various deployment types, providing several lower-level transport mechanisms to run \ac{NDN} over UDP, TCP, Ethernet, WebSocket, and others. In addition, \ac{NFD} can be easily extended to provide new transport mechanisms compatible with other underlying technologies. Despite this, the official \ac{NDN} simulator, ndnSIM 2~\cite{Mastorakis2017a}, provides \ac{NFD} only with transport to NetDevice (abstraction of link technology), thus, supporting only native \ac{NDN} deployments. To support \ac{NDN} over IP in ndnSIM 2, we have implemented a new transport mechanism and described it in Section~\ref{sec:experimental_setup}.

\section{Related Work}\label{sec:related_work}
We are not aware of any study that performs a comparative evaluation between native and overlay deployments of \ac{NDN} networks. Hence, this study is of paramount importance for further investigation of \ac{NDN} performance over wireless networks. Furthermore, simulation-based studies commonly work with native \ac{NDN}, while studies based on testbed experiments usually work with overlay \ac{NDN}. Thus, we believe that this work is the first effort to compare these \ac{NDN} deployments.

The standard \ac{NDN} architecture has no layer-2 address resolution. Thus, the native \ac{NDN} only transmits through broadcast communication, which requires several precautions to avoid storms in wireless networks, since a broadcast transmission uses only the basic service~\cite{IEEE802.11-2021} provided by most 802.11 variants, where throughput is much lower and retransmissions are disabled, providing less reliability. Conversely, the overlay \ac{NDN}, when running over the IP, can use the existing layer-2 address resolution mechanism. Thus, overlay \ac{NDN} has name-based routing, with hop-by-hop communication, but without the need for all transmissions to occur through broadcast.

Some works have proposed mechanisms to make broadcast transmissions more responsive on the wireless channel. Those researches~\cite{Li2015a,Wu2018,Shi2017,Liang2020} neither consider IoV scenarios nor comparisons between native and overlay deployments of \ac{NDN}. An approach called NLB~\cite{Li2015a} has been proposed for efficient live video broadcasting over overlay NDN in wireless networks. NLB is a leader-based mechanism to suppress duplicate requests, where a single consumer requests (via UDP unicast) and everyone receives the same data (via UDP broadcast).

A multicast rate adaptation scheme in wireless networks has been proposed in~\cite{Wu2018}. With this approach, interests are always sent via layer-2 broadcast. However, a mapping mechanism between the \ac{PIT} entry and the layer-2 address has been developed that allows the sending of data via layer-2 unicast. In this way, the proposed scheme can decide when it is better to send data packets via unicast or broadcast.

A broadcast-based adaptive forwarding strategy called self-learning has been proposed~\cite{Shi2017} and improved~\cite{Liang2020} to learn paths without needing routing algorithms. This is useful in wireless networks where nodes can be mobile and routes can change dynamically. To learn routes, the strategy broadcasts the first interest and upon receiving the data, it learns which paths have the content with the respective prefix. This way, the next interests can be sent via unicast to the learned paths. However, this approach does not perform layer-2 address mapping, and it is not possible to perform unicast with native NDN. The experiment in~\cite{Liang2020} were performed with overlay NDN, using UDP unicast and UDP broadcast.

In the specific context of \ac{VNDN},~\cite{Grassi2014} performed a study in which all packets are sent via layer-2 broadcast. However, to reduce the disadvantages of exhaustive broadcast, the authors created a mechanism that uses \ac{GPS} information to perform forwarding based on distance, avoiding two nearby cars from sending packets simultaneously to use the wireless channel more efficiently. Moreover, to restrict the spread of interest packets, a hop limitation has been applied.

The dynamic unicast~\cite{Anastasiades2016} is a routing protocol devised to perform an implicit content discovery through broadcast transmissions and dynamic content retrieval with efficient unicast links, without the need for location information. When a unicast path is broken, it can be reestablished when new interests are sent via broadcast by neighboring nodes.

Another protocol, called LOCOS~\cite{Coutinho2018}, has been proposed for content discovery and retrieval in \ac{VNDN}. LOCOS performs a directed search for content based on the location. Once the producer changes their location, requests cannot be satisfied until the new location is discovered. The protocol will periodically conduct a controlled search in the vicinity area to find the new location through transmissions of interests via broadcast. In this way, LOCOS reduces the storm problem while forwarding is directed to the nearest source.

MobiVNDN~\cite{Duarte2019} is a variant of the \ac{NDN} for \ac{VNDN} and has been proposed to mitigate the performance problems of \ac{VNDN} in wireless networks. In this proposal, the interest and data packets have some differences from the standard \ac{NDN}. Moreover, a new packet called advertisement has been proposed to propagate content availability. In MobiVNDN, vehicles exchange location and speed information with each other to assist in forwarding and calculating the probability of communication interruptions. In this approach, the geographical location provided by the GPS also performs a key role in preventing unnecessary use of the wireless channel and thus minimizing the problems of broadcast storms. Still, even though MobiVNDN makes better use of the wireless channel, communication is also done through broadcast at layer 2.

An approach has been proposed~\cite{Wang2020} to improve data delivery on \ac{VNDN} with a scheme in which the vehicular backbone has a unicast data delivery process. Despite a small scenario with few vehicular nodes, the simulation results show an increase in efficiency and the authors conclude that unicast is one of the responsible for reducing communication costs in wireless networks.

In summary, while some works focus on improving \ac{VNDN} by suppressing redundant transmissions of interest packets based on the location to minimize the broadcast storm problem, other address the problem by creating strategies where communication changes from broadcast to unicast. In this work, we take another approach by not modifying any standard \ac{NDN} behavior. We evaluate and compare the vanilla \ac{NDN} in two different deployments, the native and the overlay.

\section{Experimental Setup}\label{sec:experimental_setup}
\begin{figure}[htbp]
	\centering
	\subfloat[Native \ac{NDN}, default simulator deployment.\label{sfig:native_ndn_deployment}]{%
		\includegraphics[width=0.28\columnwidth]{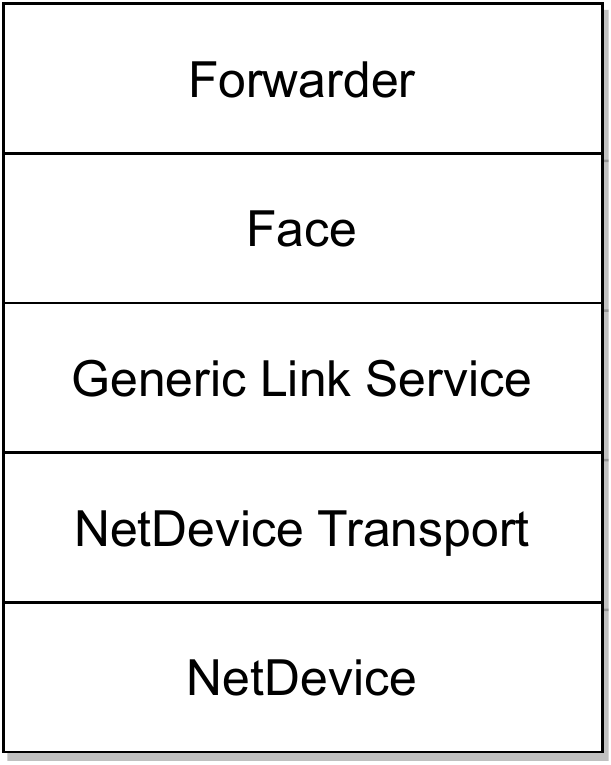}%
	}\hspace*{15pt}
	\subfloat[Overlay \ac{NDN} on UDP/IP, deployment of this work.\label{sfig:udp_overlay_ndn_deployment}]{%
		\includegraphics[width=0.28\columnwidth]{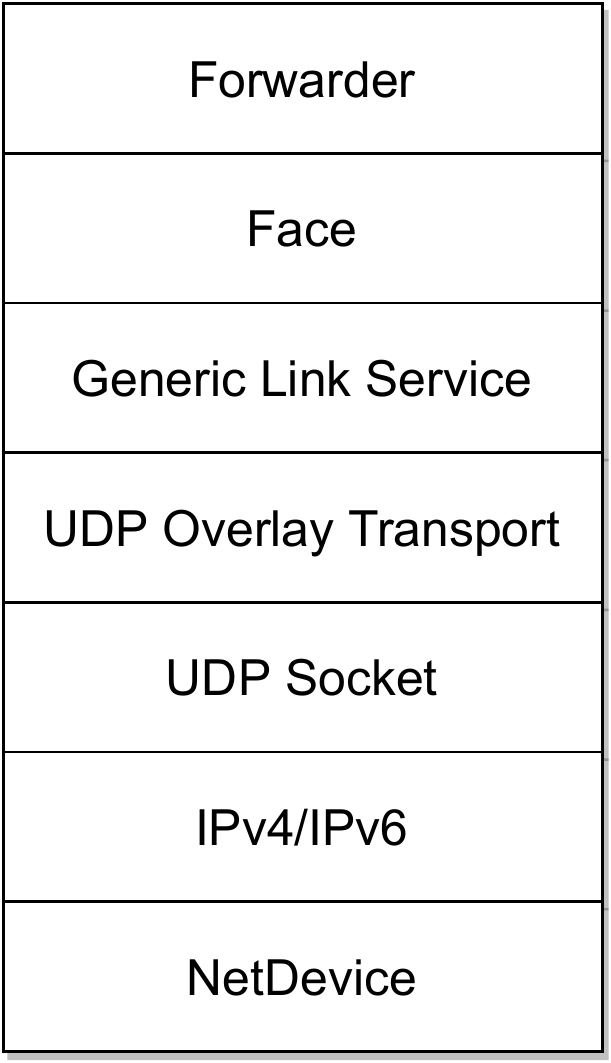}%
	}
	\caption{Overview of the \ac{NDN} deployments in the network simulator.}
	\label{fig:ndn_deployment}
\end{figure}

We performed our experiments with ndnSIM 2.8~\cite{Mastorakis2017a}, adding the \ac{NDN} stack to a modified ns-3~\cite{ns3}. However, the version used is still 3.30.1, so we migrated to ns-3.33 due to the new features of the 802.11ax module. Our experiments used two \ac{NDN} deployments (see Figure~\ref{fig:ndn_deployment}) over Wi-Fi 6 networks in the vehicular context.

Figure~\ref{sfig:native_ndn_deployment} shows the overview of native \ac{NDN}, which is the default deployment in ndnSIM 2. On the top is the forwarder that creates and manages faces (interface generalization). In the end nodes, the forwarder intermediates the communication between application and face. Each face is composed of link service and transport mechanism, the former translates the packets to lower layers, while the latter deals with the underlying communication~\cite{Afanasyev2018a}. Finally, NetDevice is an abstraction in the ns-3 of the network interface, which can be Ethernet, Wi-Fi, point-to-point, among other link technologies.

The forwarder included in ndnSIM 2, known as \ac{NFD}~\cite{Afanasyev2018a}, does not implement any overlay \ac{NDN} approach into the simulator. Thus, we developed our implementation. Figure~\ref{sfig:udp_overlay_ndn_deployment} shows the overview of our deployment of the overlay \ac{NDN} on top of IP network. The main change is that the transport layer of each face creates an unicast tunnel to a remote face through UDP socket over IPv4 or IPv6. These tunnels are created hop-by-hop between each NDN node. Therefore, in this deployment, a UDP socket is the underlying communication mechanism for the \ac{NDN} network. Finally, for this deployment to work properly in ndnSIM 2, we need to modify the stack helper and global routing helper to configure the communication via UDP sockets automatically.

\subsection{Vehicular Traffic Modeling}\label{sec:traffic_modeling}
\begin{figure}[htbp]
	\centerline{\includegraphics[width=0.8\columnwidth]{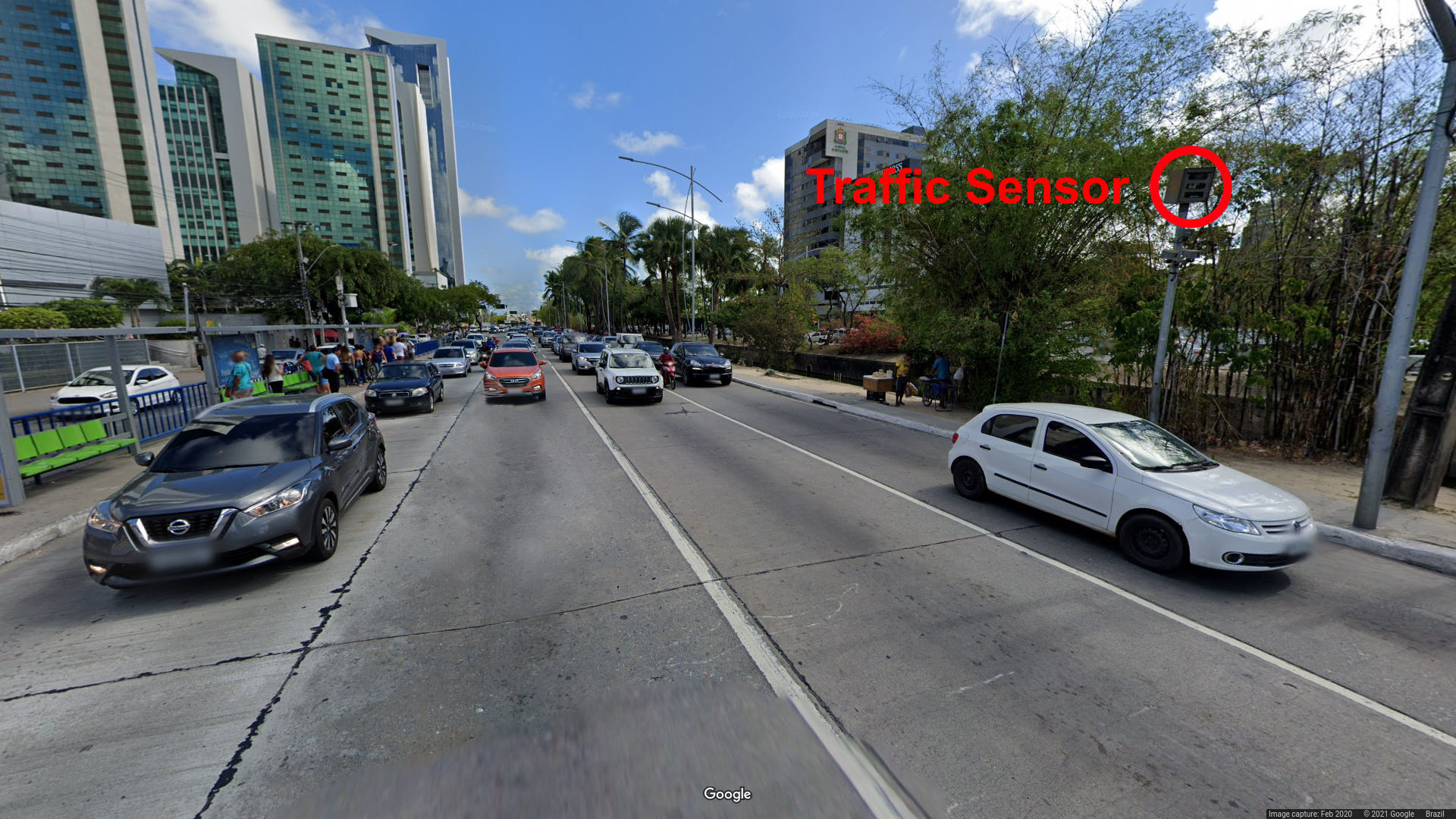}}
	\caption{Avenue with four lanes modeled on SUMO.}
	\label{fig:agamenon}
\end{figure}

To model vehicular traffic realistically in \ac{SUMO} \cite{Lopez2018}, we collected open data~\cite{cttu2019recife} from the transportation authority of Recife, Brazil, and we chose the data of 2019, as this year vehicular traffic was not influenced by the Covid-19 pandemic. The transport authority provides data such as the date/time and speed of each car traveling the streets for all city traffic sensors. The traffic sensor identified by FS037REC was chosen to have its data analyzed. As shown in Figure~\ref{fig:agamenon}, it monitors one of the main avenues in the city. 

We calculate the average traffic on business days and based on this, we model the scenario with 172 meters of avenue, 3 bus stops and 125 vehicles over 300 seconds, with an average and a maximum speed of 31 km/h and 60 km/h, respectively. Finally, we performed 31 simulations of this modeling to import the flow generated in each of the instances of the network simulations (see Table~\ref{tab:list_evaluated_scenarios}) and calculate the statistical tests.

\subsection{Scenarios}\label{sec:scenario}
\begin{figure}[htbp]
	\centerline{\includegraphics[width=0.8\columnwidth]{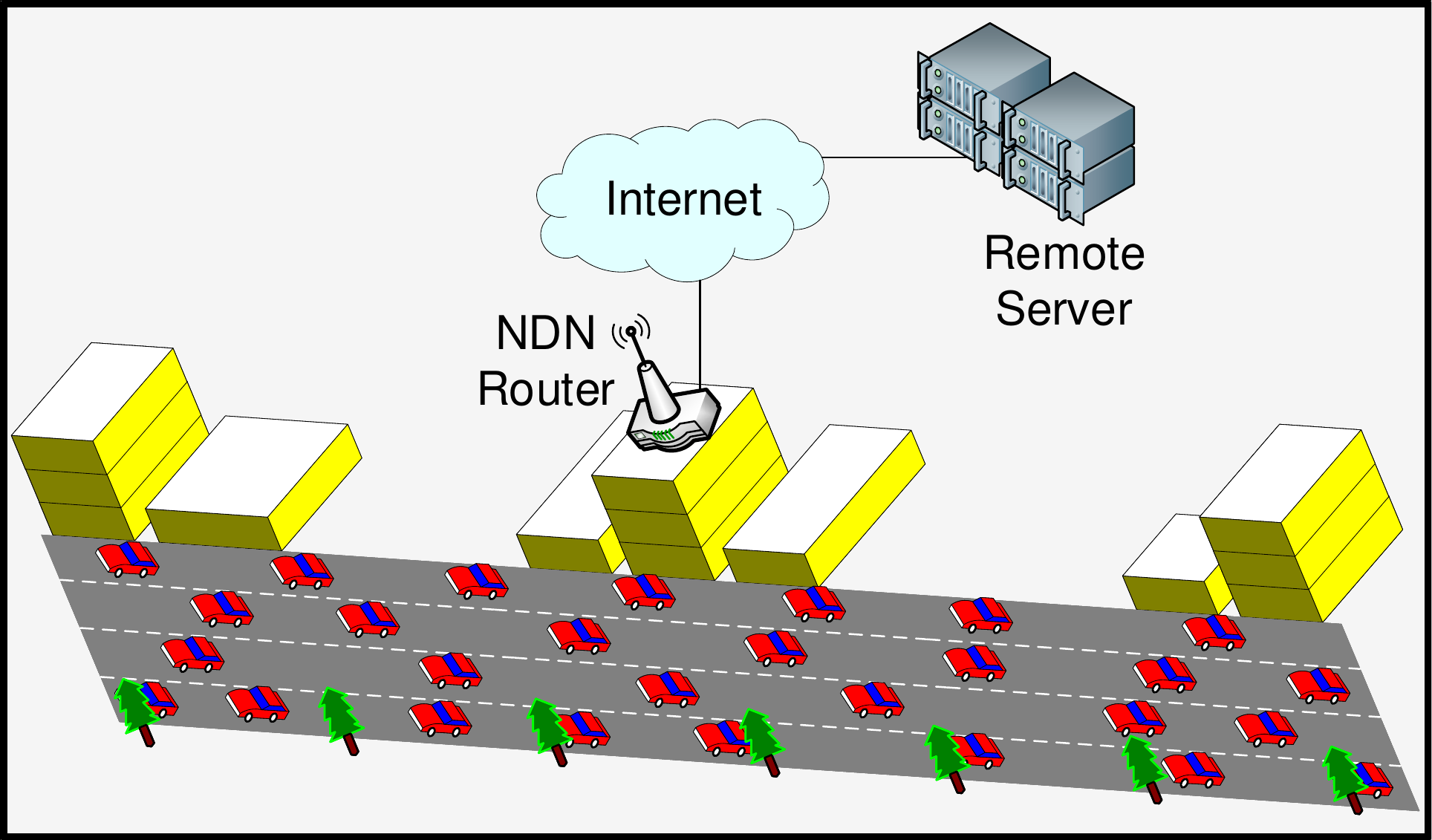}}
	\caption{Vehicular NDN network topology.}
	\label{fig:topology}
\end{figure}

Our simulation scenarios consist of 125 vehicular nodes that, along the 172 meters of the avenue, will be connected to an \ac{NDN} router through a Wi-Fi 6 network, 802.11ax standard with \ac{MCS}~11 and 800ns of \ac{GI}. As shown in Figure~\ref{fig:topology}, the \ac{NDN} router has been placed halfway. In addition, it is connected to the remote server (the producer) with a 1~Gbps point-to-point link and 30~ms delay. The \ac{CS} size of the \ac{NDN} router is 10,000 packets and for all other nodes it is 0. The payload of the data packets is 1024 bytes. Finally, when using the overlay \ac{NDN} deployment, we configure the networks with IPv4 addressing.

In our scenarios, the vehicular nodes send interest packets at a constant rate, defined uniformly between 50 and 100 packets per second for each vehicle. We created two scenarios identified by a suffix 1 or 2 at the end of the name for each instance of deployment (please, see Table~\ref{tab:list_evaluated_scenarios}). In the first, all vehicular nodes use the \textit{ConsumerCbr} application, available by default in the ndnSIM 2. We define that each vehicle requests content with a different prefix, so we force all vehicles to request distinct data between them. Because of this peculiarity, there should be no advantage to broadcast traffic. In the second scenario, we randomly choose 50\% of the vehicles to use the \textit{ConsumerCbr} application in the same way as in the first scenario, and the rest to use the \textit{ModifiedConsumerCbr}, a new application modified by us that sends interest packets with the sequence number based on the simulation time, therefore, vehicles request contents with the same name at the same time. Consequently, many vehicles request the same content and there may be an advantage in broadcast traffic. In these proposed scenarios, we performed 31 simulations for each instance present in Table~\ref{tab:list_evaluated_scenarios}.

\begin{table}[htbp]
	\centering
	\caption{List of evaluated instances}
	\begin{tabular}{ccc}
		\hline NDN Deployment & Scenarios & Instance\\ 
		\hline
		\hline \multirow{2}{*}{Native} & 1 & Native-1  \\
		\cline{2-3} & 2 & Native-2 \\
		\hline \multirow{2}{*}{Overlay} & 1 & Overlay-1   \\
		\cline{2-3} & 2 & Overlay-2   \\
		\hline
	\end{tabular}
	\label{tab:list_evaluated_scenarios}
\end{table}

\section{Performance Evaluation}\label{sec:performance_evaluation}
In this section, we inform the statistical methods used, as well as present and discuss the results obtained with the simulations performed.

\subsection{Statistical Tests}\label{sec:statistical_tests}

Arcuri and Briand~\cite{Arcuri2011} discuss the usage of statistical testing to analyze randomized algorithms in software engineering. Based on that, we chose Shapiro-Wilk to test the normality of the results. Although the data follow a normal distribution, homoscedasticity is not satisfied, that is, the variances between the distributions are not equivalent. Thus, we chose to use the following statistical tests: Mann-Whitney U-test, a non-parametric significance test; Vargha and Delaney's $\hat{A}_{12}$, a non-parametric effect size test, for assessing whether there are statistical differences among the obtained results. We used a confidence level of $95\%$ in all cases. All statistical analyses and tests were run using SciPy~\cite{2020SciPy-NMeth}, an open-source scientific tool.

\subsection{Results}\label{sec:results}
We started the discussion with the Mann-Whitney U-test, that deals with their stochastic ranking~\cite{Arcuri2011} to observe the probability that one population will have its values higher than the other and thus verify the statistical significance between these populations. Our null hypothesis ($H_{0}$) is not rejected when the p-value is greater than 0.05, suggesting that the evaluated instances achieved the same performance, otherwise, our alternative hypothesis ($H_{a}$) suggests that the instances achieved statistically different performances.

\begin{figure*}[tb]
	\centering
	\subfloat[The Mann-Whitney U-test.\label{fig:mw}]{%
		\includegraphics[width=0.5\columnwidth]{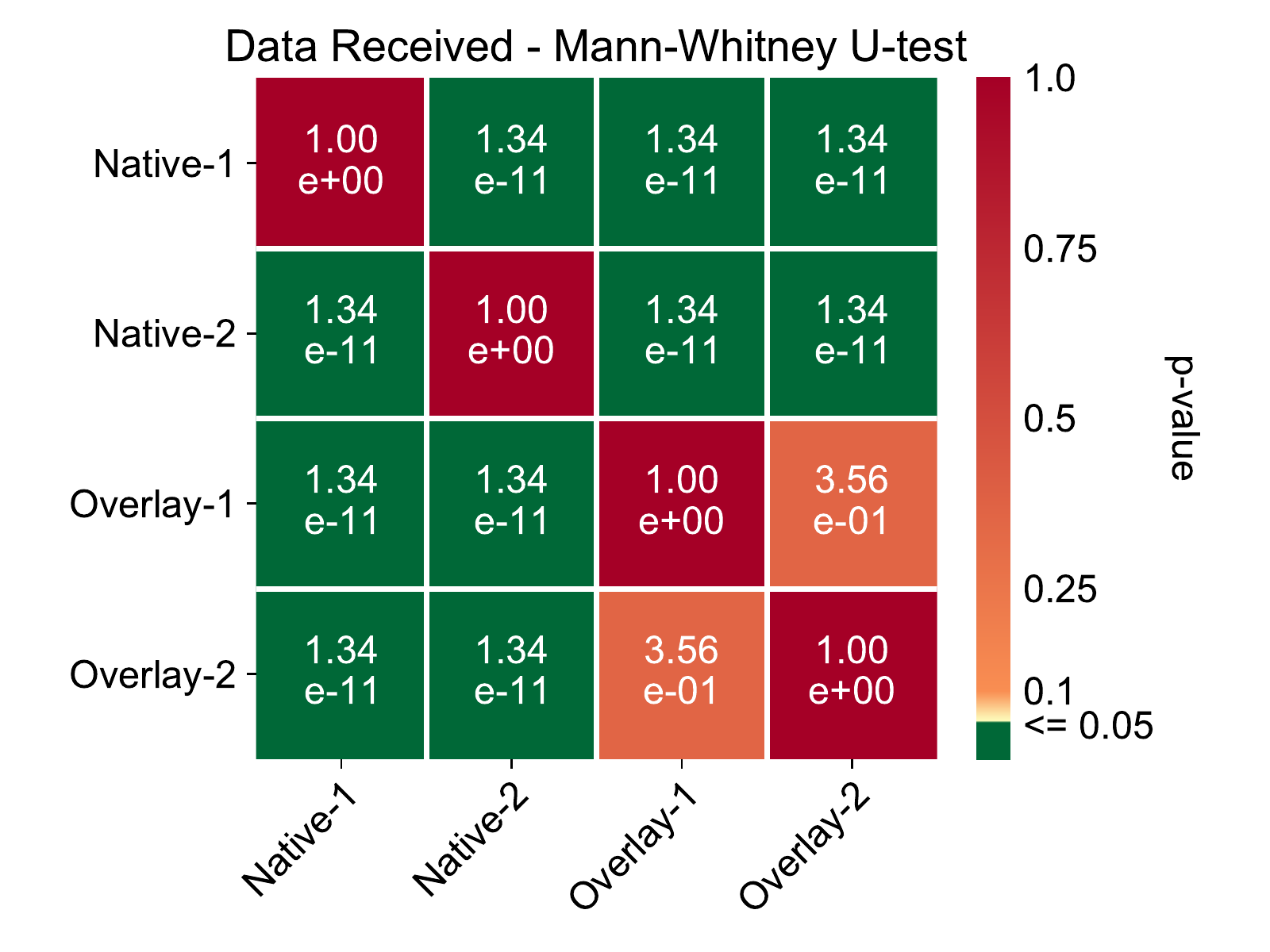}%
	}
	\subfloat[The Vargha and Delaney's $\hat{A}_{12}$ index.\label{fig:vd}]{%
		\includegraphics[width=0.5\columnwidth]{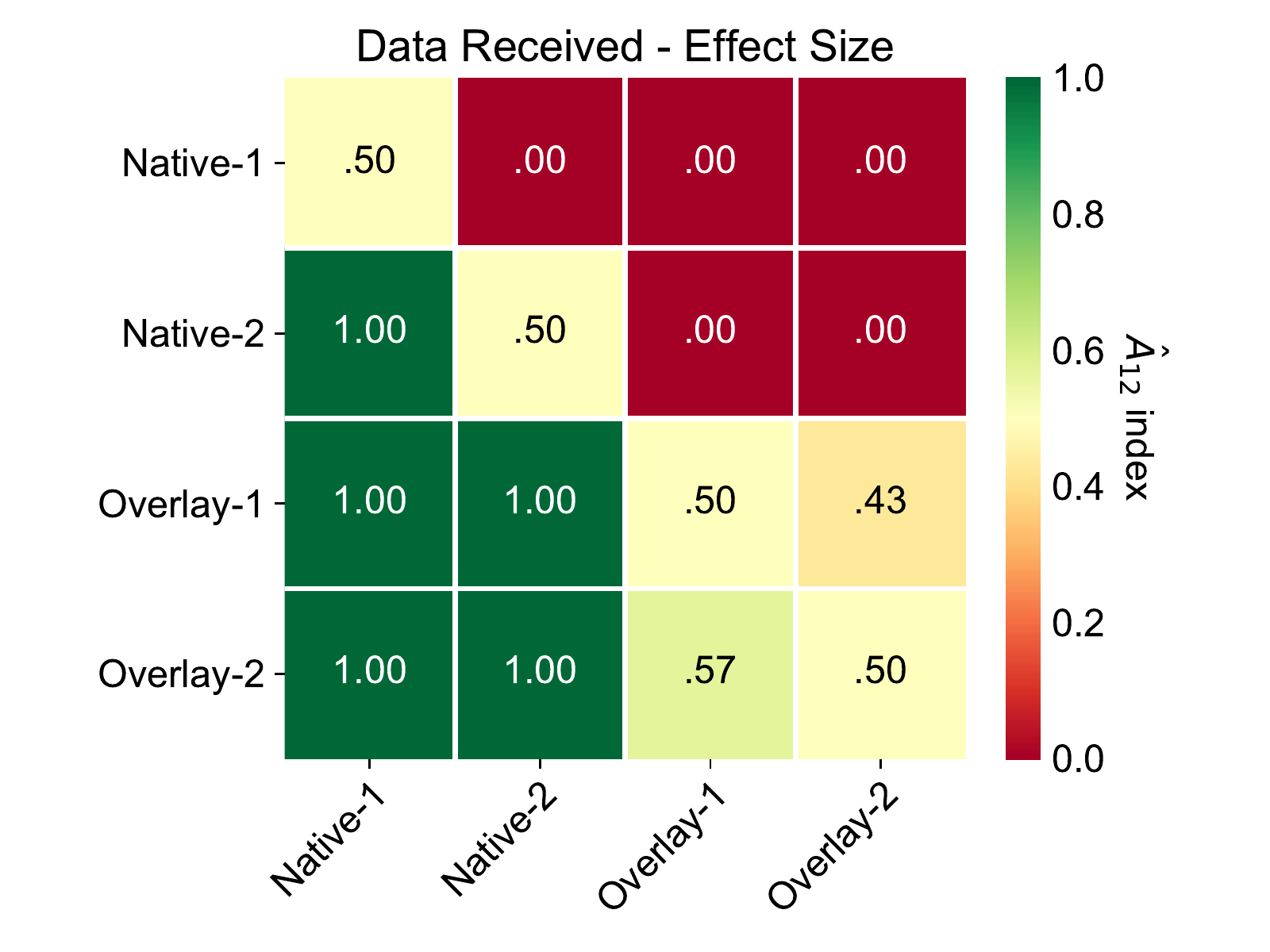}%
	}
	\caption{Statistical tests of significance and effect size.}
	\label{fig:statistical_tests}
\end{figure*}

Figure~\ref{fig:mw} shows U-test p-values for the metric of data received in vehicles in each instance evaluated (see Table~\ref{tab:list_evaluated_scenarios}). The only time that the $H_{0}$ was not rejected was when we compared the overlay \ac{NDN} network in the two scenarios. It occurred because the transmissions for this \ac{NDN} deployment are face-to-face unicast, and due to the layer-2 address resolution traditionally provided by the IP stack was possible to have full access to wireless network services. Thus, there was no performance difference of overlay deployment in the two evaluated scenarios. It is also important to note that the $H_{a}$ was accepted when comparing the native \ac{NDN} deployment in the two evaluated scenarios. Since this deployment performs only broadcast transmissions, it is expected to have a performance difference in the scenarios evaluated, even having access only to basic service~\cite{IEEE802.11-2021} of wireless network, due to the exhaustive broadcast.

We use the $\hat{A}_{12}$ effect size test to analyze also the magnitude of the difference. This test presents an intuitive result, measuring the probability that one approach is better than another. Figure~\ref{fig:vd} shows $\hat{A}_{12}$ index for the metric of data received. The native deployment presented the worst $\hat{A}_{12}$ index, mainly in the first scenario. According to the result of this test, it is unlikely that the native deployment will perform better than the overlay deployment. When looking at Native-1 versus Native-2 instances, it is clear that if consumers request the same content for this deployment, it will make a significant difference, improving performance. When observing the Overlay-1 and Overlay-2 instances, the $\hat{A}_{12}$ index for the second presented a slight superiority despite the U-test showing no statistical difference. The slight difference could be explained because much of the data requested is already in the \ac{NDN} router's \ac{CS}. Thus, other vehicles have already requested data, something that does not occur in the first scenario.

\begin{figure*}[tb]
	\centering
	\subfloat[Total values.\label{fig:total_values_bar}]{%
		\includegraphics[width=0.5\columnwidth]{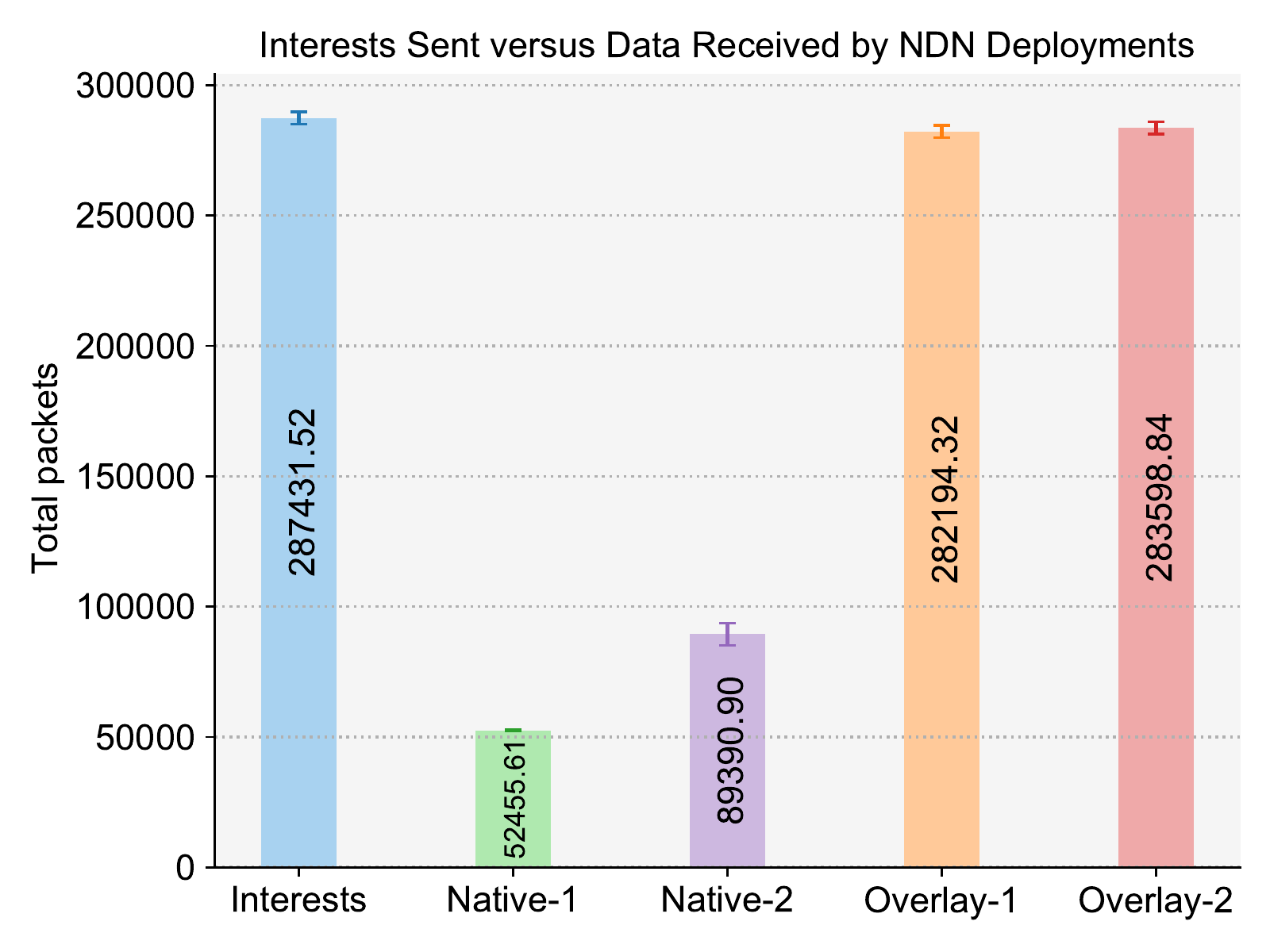}%
	}
	\subfloat[Values in relation to time.\label{fig:values_in_time_line}]{%
		\includegraphics[width=0.5\columnwidth]{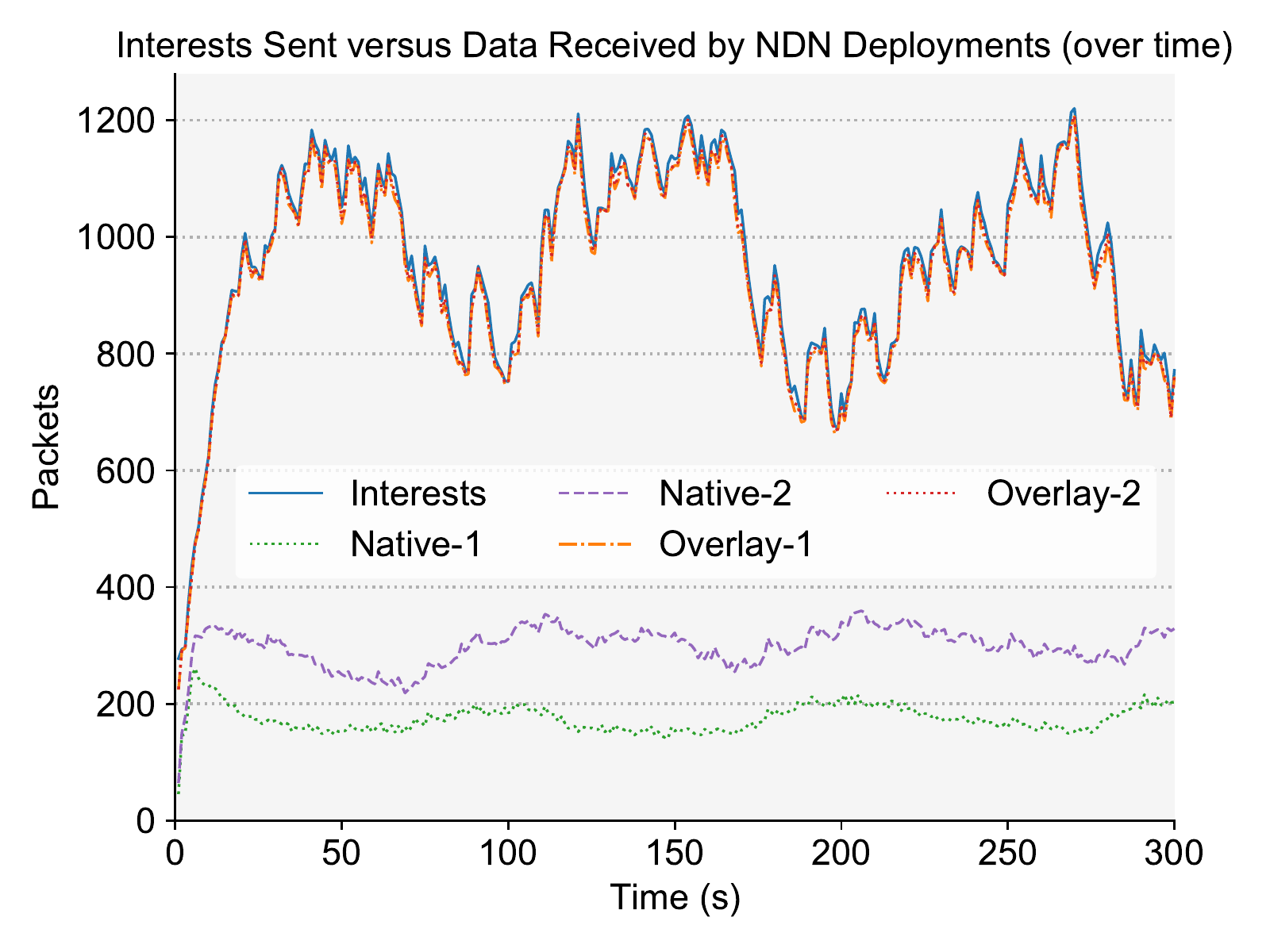}%
	}\\
	\subfloat[Total values by Application.\label{fig:total_values_by_app}]{%
		\includegraphics[width=0.5\columnwidth]{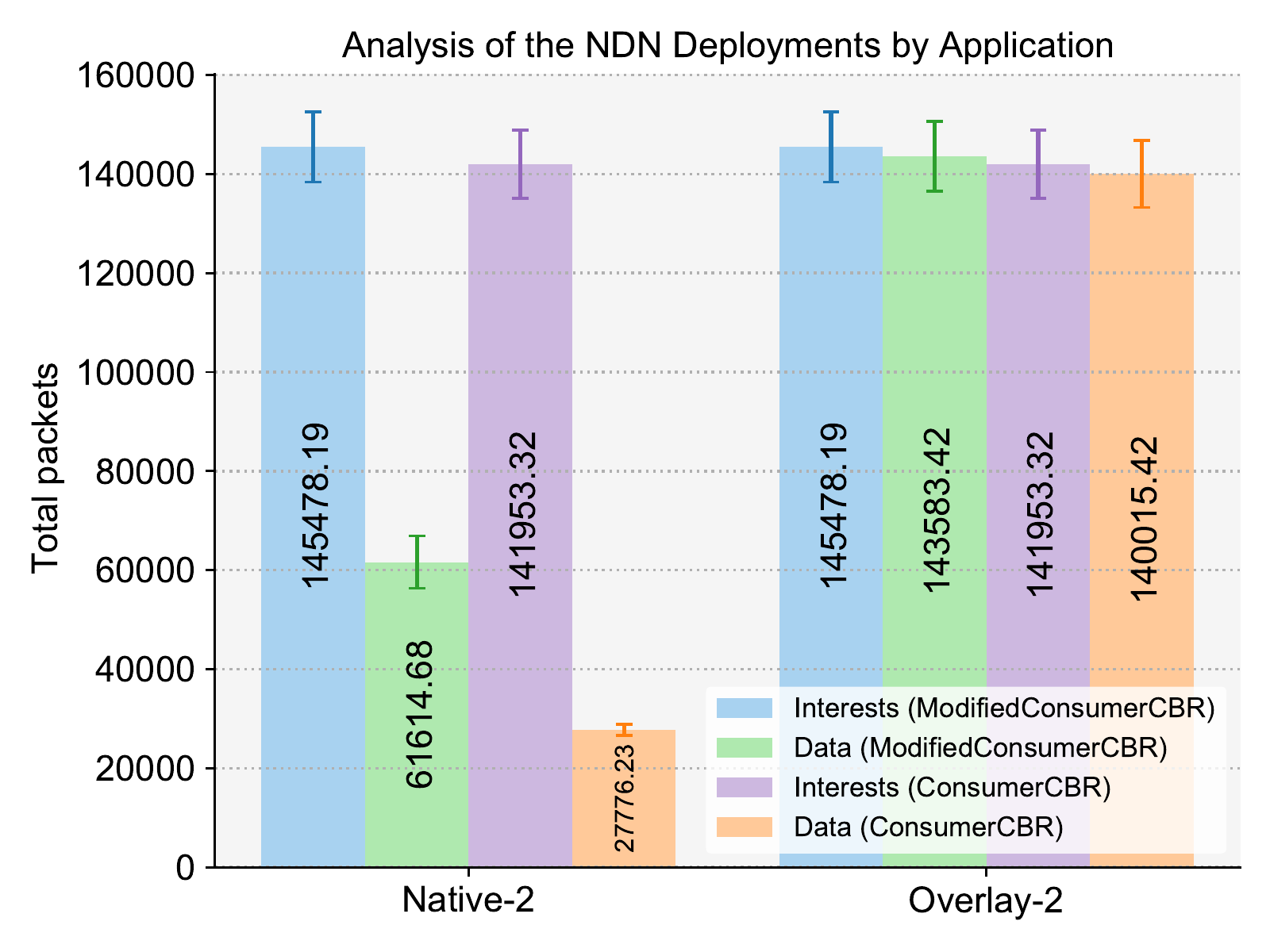}%
	}
	\caption{Relation between interest sent and data received by evaluated instances.}
	\label{fig:interest_sent_versus_data_received}
\end{figure*}

Figure~\ref{fig:total_values_bar} shows the total number of interest packets sent and the data received in each evaluated instance. The number of interest packets sent is the same, regardless of the instance. Therefore, only one bar has been placed on the chart with this information. It is possible to observe that there is a considerable superiority of Native-2 over Native-1, around 70\% of more data received, with Native-1 reaching only 18.25\% of requests satisfied, against 31.1\% reached by Native-2. Thus, it confirms the importance of vehicles requesting the same content in the native deployment. However, when comparing Native-2 with the two instances of the overlay deployment, the performance difference is significant, receiving more than 200\% of data received than Native-2, reaching values close to 99\% of requests satisfied in overlay deployments.

This difference is explained by the fact that the wireless network standard offers a basic service of communication for broadcast transmissions~\cite{IEEE802.11-2021}, as a consequence, the performance of this traffic is reduced. The Figure~\ref{fig:values_in_time_line} shows the same information, but from a new perspective to show the packets in relation to the simulation time. The lines of data received in the overlay instances always follow the interests sent, taking advantage of wireless network resources, unlike the instances of native deployment that always use the basic service of the wireless network, falling away from optimal performance.

Figure~\ref{fig:total_values_by_app} also shows this analysis of the relationship between interests and data, but by application, which is why it contains instances only from the second scenario. In our experiments the vehicles running \textit{ConsumerCbr} requests distinct data, while the vehicles running \textit{ModifiedConsumerCbr} requests same data. Hence, we compared NDN deployments with these two types of traffic. In the native deployment, the \textit{ConsumerCbr} application had only 19.57\% of requests satisfied and \textit{ModifiedConsumerCbr} increased only 42.35\%. This shows that even when the same data are requested, native NDN has difficulties in performing satisfactory use of the available resources on wireless networks. Conversely, in the overlay deployment, both applications reached values close to 99\%.

All of these results showed that the excessive use of broadcast transmissions in the link layer by native NDN is not scalable and reduces throughput in wireless networks. Therefore, it is essential to develop a layer-2 address mapping mechanism to native \ac{NDN}.

\section{Conclusions}\label{sec:conclusions}
In this paper, we implemented the overlay \ac{NDN} over IP in ndnSIM 2.8 simulator and we conducted a comparative evaluation with the native \ac{NDN} deployment in vehicular network with Wi-Fi 6. Our vehicular traffic has been based on real traces, and from this, we propose two scenarios. In the first, all vehicles request distinct data, while in the second scenario half of the vehicles request the same data. 

We evaluate which deployment achieves the best rate of satisfied requests, since the native deployment only performs broadcast transmission. Oppositely, the overlay deployment in our scenarios only performs unicast transmission. Our results show that the native deployment has low rates of satisfied requests in Wi-Fi 6 due to the extensive use of broadcast transmissions, achieved only 42.35\% in the best case. On the other hand, the overlay deployment reached values close to 99\%.

\ac{NDN} is a clean-slate network architecture. It is unlikely to be initially deployed in the native form. A transition period will be necessary for this migration to take place smoothly. Thus, comprehensively understanding the performance of the overlay \ac{NDN} deployment is a necessary step for its adoption in vehicular networks. As future work, a layer-2 address mapping mechanism should be devised to improve the performance of native NDN deployment in wireless networks and compared with existing solutions. In addition, more scenarios should be investigated considering heterogeneous nodes with native \ac{NDN}, overlay \ac{NDN}, and nodes that have only IP stack, all communicating with each other.

\subsubsection*{Acknowledgments.}

This work was partially supported by the National Council for Scientific and Technological Development (CNPq) (Grant No. 312831/2020-0).

%
%
%
\bibliographystyle{splncs04}
\bibliography{mybibliography}

\begin{thebibliography}{10}
\providecommand{\url}[1]{\texttt{#1}}
\providecommand{\urlprefix}{URL }
\providecommand{\doi}[1]{https://doi.org/#1}

\bibitem{Afanasyev2018a}
{Afanasyev, A. \textit{et al}.}: {NFD Developer's Guide} (2018), technical
  Report NDN-0021

\bibitem{Anastasiades2016}
Anastasiades, C., Weber, J., Braun, T.: {Dynamic Unicast: Information-centric
  multi-hop routing for mobile ad-hoc networks}. Computer Networks
  \textbf{107},  208--219 (2016), mobile Wireless Networks

\bibitem{Arcuri2011}
Arcuri, A., Briand, L.: {A Practical Guide for Using Statistical Tests to
  Assess Randomized Algorithms in Software Engineering}. In: 2011 33rd
  International Conference on Software Engineering (ICSE). pp. 1--10 (2011)

\bibitem{Coutinho2018}
{Coutinho}, R.W.L., {Boukerche}, A., {Yu}, X.: {A Novel Location-Based Content
  Distribution Protocol for Vehicular Named-Data Networks}. In: 2018 IEEE
  Symposium on Computers and Communications (ISCC). pp. 01007--01012 (June
  2018)

\bibitem{cttu2019recife}
CTTU: {Open Data of Vehicle Traffic from Recife-Brazil} (2019),
  \url{http://dados.recife.pe.gov.br/dataset/velocidade-das-vias-quantitativo-por-velocidade-media-2019}

\bibitem{Duarte2019}
Duarte, J.M., Braun, T., Villas, L.A.: {MobiVNDN: A distributed framework to
  support mobility in vehicular named-data networking}. {Ad Hoc Networks}
  \textbf{82},  77--90 (2019)

\bibitem{Fahrianto2021a}
Fahrianto, F., Kamiyama, N.: {A Low-Cost IP-to-NDN Translation Gateway}. In:
  2021 IEEE 22nd International Conference on High Performance Switching and
  Routing (HPSR). pp.~1--5 (June 2021)

\bibitem{Grassi2014}
{Grassi}, G., {Pesavento}, D., {Pau}, G., {Vuyyuru}, R., {Wakikawa}, R.,
  {Zhang}, L.: {VANET via Named Data Networking}. In: 2014 IEEE Conference on
  Computer Communications Workshops (INFOCOM WKSHPS). pp. 410--415 (April 2014)

\bibitem{Khorov2019}
Khorov, E., Kiryanov, A., Lyakhov, A., Bianchi, G.: {A Tutorial on IEEE
  802.11ax High Efficiency WLANs}. {IEEE Communications Surveys Tutorials}
  \textbf{21}(1),  197--216 (2019)

\bibitem{Li2015a}
Li, M., Pei, D., Zhang, X., Zhang, B., Xu, K.: {NDN Live Video Broadcasting
  over Wireless LAN}. In: 2015 24th International Conference on Computer
  Communication and Networks (ICCCN). pp.~1--7 (Aug 2015)

\bibitem{Liang2020}
Liang, T., Pan, J., Rahman, M.A., Shi, J., Pesavento, D., Afanasyev, A., Zhang,
  B.: {Enabling Named Data Networking Forwarder to Work Out-of-the-Box at Edge
  Networks}. In: 2020 IEEE International Conference on Communications Workshops
  (ICC Workshops). pp.~1--6 (2020)

\bibitem{Lopez2018}
{Lopez, P. A, \textit{et al}.}: {Microscopic Traffic Simulation using SUMO}.
  In: {2018 21st International Conference on Intelligent Transportation Systems
  (ITSC)}. pp. 2575--2582 (2018)

\bibitem{Mastorakis2017a}
Mastorakis, S., Afanasyev, A., Zhang, L.: {On the Evolution of ndnSIM: an
  Open-Source Simulator for NDN Experimentation}. SIGCOMM Comput. Commun. Rev.
  \textbf{47}(3),  19--33 (Sep 2017)

\bibitem{Nour2019}
Nour, B., Li, F., Khelifi, H., Moungla, H., Ksentini, A.: {Coexistence of ICN
  and IP Networks: An NFV as a Service Approach}. In: 2019 IEEE Global
  Communications Conference (GLOBECOM). pp.~1--6 (Dec 2019)

\bibitem{ns3}
{ns-3}: {ns-3 Network Simulator Website} (2021), \url{https://www.nsnam.org/}

\bibitem{IEEE802.11-2021}
SA, I.: {IEEE Standard for Information Technology -- Telecommunications and
  Information Exchange between Systems - Local and Metropolitan Area
  Networks--Specific Requirements - Part 11: Wireless LAN Medium Access Control
  (MAC) and Physical Layer (PHY) Specifications}. {IEEE Std 802.11-2020
  (Revision of IEEE Std 802.11-2016)} pp. 1--4379 (Feb 2021)

\bibitem{Shi2017}
Shi, J., Newberry, E., Zhang, B.: {On Broadcast-based Self-Learning in Named
  Data Networking}. In: 2017 IFIP Networking Conference (IFIP Networking) and
  Workshops. pp.~1--9 (June 2017)

\bibitem{2020SciPy-NMeth}
{Virtanen, P. \textit{et al}.}: {{SciPy} 1.0: Fundamental Algorithms for
  Scientific Computing in Python}. {Nature Methods}  \textbf{17},  261--272
  (2020)

\bibitem{Wang2020}
Wang, X., Wang, Z., Cai, S.: {Data Delivery in Vehicular Named Data
  Networking}. IEEE Networking Letters  \textbf{2}(3),  120--123 (2020)

\bibitem{Wu2018}
Wu, F., Yang, W., Fan, Z., Tian, K.: {Multicast Rate Adaptation in WLAN via
  NDN}. In: 2018 27th International Conference on Computer Communication and
  Networks (ICCCN). pp.~1--8 (July 2018)

\bibitem{Zhang2014}
Zhang, L., Afanasyev, A., Burke, J., Jacobson, V., claffy, k., Crowley, P.,
  Papadopoulos, C., Wang, L., Zhang, B.: {Named Data Networking}. SIGCOMM
  Comput. Commun. Rev.  \textbf{44}(3),  66--73 (Jul 2014)

\end{thebibliography}
\end{document}